\documentclass[onecolumn,floatfix,showpacs,showkeys,preprintnumbers,nofootinbib,superscriptaddress]{revtex4}
\usepackage[utf8]{inputenc}
\usepackage[sort&compress]{natbib}
\usepackage{ulem}
\usepackage{bm}
\usepackage{times}
\usepackage{amssymb,amsbsy,amsmath,amsfonts}
\usepackage{graphicx}
\usepackage{float}
\usepackage{color}
\usepackage{morefloats}
\usepackage{rotating}
\usepackage{srcltx}
\usepackage{slashed}
\usepackage{subfigure}
\usepackage{multirow}
\usepackage{verbatim}
\usepackage{hyperref}
\usepackage{tabularx}
\usepackage{adjustbox}

\usepackage{overpic}
\usepackage{makecell}

\begin{document}

\title{ Productions of  $T_{cc}$ and its SU(3)-flavor symmetry and heavy quark spin symmetry partners  in   $B_c$ decays   }

\author{Yi Zhang}
\affiliation{School of Physics,  Beihang University, Beijing 102206, China}

\author{Ming-Zhu Liu}
\email[Corresponding author: ]{liumz@lzu.edu.cn}
\affiliation{
Frontiers Science Center for Rare isotopes, Lanzhou University,
Lanzhou 730000, China}
\affiliation{ School of Nuclear Science and Technology, Lanzhou University, Lanzhou 730000, China}

\author{Li-Sheng Geng}\email[Corresponding author: ]{lisheng.geng@buaa.edu.cn}
\affiliation{School of Physics,  Beihang University, Beijing 102206, China}
\affiliation{Sino-French Carbon Neutrality Research Center, \'Ecole Centrale de P\'ekin/School of General Engineering, Beihang University, Beijing 100191, China}
\affiliation{Peng Huanwu Collaborative Center for Research and Education, Beihang University, Beijing 100191, China}
\affiliation{Beijing Key Laboratory of Advanced Nuclear Materials and Physics, Beihang University, Beijing 102206, China }
\affiliation{Southern Center for Nuclear-Science Theory (SCNT), Institute of Modern Physics, Chinese Academy of Sciences, Huizhou 516000, China}

\date{\today}
\begin{abstract}

Inspired by the observation of the doubly charmed tetraquark state $T_{cc}$ at $pp$ collisions in the inclusive processes, we systematically investigate the production of doubly charmed tetraquark states in exclusive  $B_c$ decays.  
In this work, we assume the $T_{cc}$ as a $DD^*$ bound state, and then predict the masses  of its heavy quark spin symmetry partner $D^*D^*$ (denoted by $T_{cc}^{*}$)  as well as their SU(3)-flavor symmetry partners, i.g.,  $D_sD^*/D_s^*D$ (denoted by $T_{ccs}^{+}$ and $T_{ccs}^{++}$) and $D^*D_s^*$ (denoted by $T_{ccs}^{*+}$ and $T_{ccs}^{*++}$), using the contact range effective field theory. Within the molecular picture, we compute their partial decays and production rates in $B_c$ meson decays. We identify the decays  $B_c  \to D^0D^0\pi^+\bar{D}^0 $ and   $B_c \to D_s^+ D^{0} \pi^+{D}^{-}$ as promising channels to observe the tetraquark states $T_{cc}^{(*)}$ and $T_{ccs}^{(*)++}$, respectively.    Finally, by combining these results with $B_c$ 
  production cross sections, we estimate the expected event yields for these states in the upcoming LHC Run 3 and 4.    Our results indicate that the $T_{ccs}^{+}$ and $T_{ccs}^{*++}$ states are likely to be observed in $B_c$ decays, while it is quite difficult for the $T_{cc}$ and $T_{cc}^{*}$ states.

\end{abstract}


\maketitle

\section{Introduction}

According to the conventional constituent quark model, hadrons are classified as mesons made of a quark and an anti-quark, and baryons made of three quarks~\cite{Gell-Mann:1964ewy}. These conventional hadrons play a vital role in advancing our understanding of the strong interaction and hadron structure. A notable example is the study of charmonium spectroscopy using the Cornell potential~\cite{Eichten:1974af}, which has successfully validated both the long-range and short-range behaviors of the strong interaction~\cite{Bali:2000gf}.
 Since 2003, an increasing number of  states beyond the conventional quark model have been discovered~\cite{Brambilla:2010cs,Olsen:2017bmm,Brambilla:2019esw}, providing new insights into hadron structure and the underlying dynamics of the strong interaction. Exotic states offer profound insights into the complexity of the strong interaction and the structure of hadronic systems. These include hadron-hadron interactions in hadronic molecules, QCD dynamics in compact multiquark states, and electromagnetic forces in hadro-charmonium systems. In addition, some exotic states are not genuine physical states but may arise from kinematical effects.
Despite extensive experimental and theoretical investigations~\cite{Chen:2016qju,Lebed:2016hpi,Oset:2016lyh,Esposito:2016noz,Dong:2017gaw,Guo:2017jvc,Ali:2017jda,Karliner:2017qhf,Guo:2019twa,Liu:2024uxn,Husken:2024rdk,Wang:2025sic,Doring:2025sgb}, the nature of these exotic states remains a subject of controversy—highlighting the intricate and challenging character of the non-perturbative strong interaction. 
 
It is well-established that hadron-hadron interactions play a pivotal role in excited or exotic hadron spectroscopy and in strong and weak decays. For the spectrum of exotic states, a prime example is the $X(3872)$~\cite{Belle:2003nnu}. A detailed analysis of its invariant mass distributions supports the interpretation that the $X(3872)$ possesses a significant molecular component~\cite{LHCb:2020xds,BESIII:2023hml,Ji:2025hjw}, which reflects the role of the $\bar{D}^*D$ interaction. Theoretical predictions for the ratio of $\mathcal{B}[\psi(2S) \to \rho \pi]$ to $\mathcal{B}[\psi(1S) \to \rho \pi]$ show significant discrepancies with experimental results, a phenomenon known as the $\rho$–$\pi$ puzzle~\cite{Franklin:1983ve}. One classical explanation is that final-state interactions(FSIs) play a vital role in addressing this long-standing puzzle~\cite{Li:1996yn,Zhao:2010mm,Zhao:2010zzv,Wang:2012mf}. FSIs play a significant role in the weak decays of heavy hadrons as well, substantially enhancing the branching fractions of nonfactorizable decay channels~\cite{Colangelo:2002mj,Jia:2024pyb,Cheng:2004ru,Ling:2021qzl}, and giving rise to significant CP violation signals in these processes~\cite{Duan:2024zjv,Wang:2024oyi}. Therefore, investigations of hadron-hadron interactions are of great value. Information from lattice QCD scattering data~\cite{Aoki:2020bew}, femtoscopy correlation functions~\cite{Liu:2024uxn}, and invariant mass distributions in exclusive and inclusive decay processes~\cite{Nakamura:2023obk,Du:2019pij} can all help improve our understanding of hadron-hadron interactions with increasing precision.

Doubly heavy tetraquark states fall into the category of exotic hadronic states. In 2021, the LHCb Collaboration first observed a doubly charmed tetraquark state $T_{cc}$ in the inclusive process at $pp$ collision, with a mass close to the $DD^*$ mass threshold~\cite{LHCb:2021vvq,LHCb:2021auc}. Unlike the case of hidden-charm tetraquarks, the $T_{cc}$ state can only be of two exotic configurations, i.e., a hadronic molecule~\cite{Dong:2021bvy,Ling:2021bir,Du:2021zzh,Feijoo:2021ppq,Albaladejo:2021vln,Xin:2021wcr,Deng:2021gnb,Fleming:2021wmk,Yan:2021wdl,Ren:2021dsi,Agaev:2022ast} or a compact tetraquark state~\cite{Agaev:2021vur,Weng:2021hje,Wu:2022gie}. While the mass and decay widths are readily explained within the molecular picture, the compact tetraquark picture has difficulty explaining the decay width. Therefore, the discovery of $T_{cc}$ motivate a lot of studies on the $DD^*$ hadron interactions in Lattice QCD~\cite{Padmanath:2022cvl,Chen:2022vpo,Lyu:2023xro,Collins:2024sfi,Whyte:2024ihh}, chiral effective field theory(ChEFT)~\cite{Wang:2022jop,Zhai:2023ejo,Xu:2025xrl}, and one-boson-exchange model~\cite{Sakai:2023syt,Asanuma:2023atv,Sun:2024wxz,Qiu:2023uno,Zhang:2024dth}. By treating the $T_{cc}^+$ as a $DD^*$ molecular state, the $DD^*$ interaction can be extracted. This interaction can then be used to investigate other phenomena related to the $DD^*$ system, such as relevant three-body hadronic molecules~$DDD^*$~\cite{Wu:2021kbu}, $\bar{\Sigma}_cDD^*$~\cite{Pan:2022xxz}, and ~$KDD^*$~\cite{Pan:2025xvq}, and heavy quark spin symmetry(HQSS) partner $D^*D^*$~\cite{Albaladejo:2021vln,Du:2021zzh,Deng:2021gnb}, SU(3)-flavor symmetry partners $D_{s}D^*$ and $D_s^*D^*$~\cite{Dai:2021vgf,Peng:2023lfw,Chen:2024bre,Liu:2023ckj}, and correlation functions~\cite{Albaladejo:2023wmv}. See, e.g., Ref.~\cite{Liu:2024uxn}. In addition, the $DD^*$ interaction can be verified in the exclusive $B_c$ decay~\cite{Li:2023hpk} and heavy ion collisions~\cite{Hu:2021gdg,Chen:2023xhd}.

In this work, we focus on the $T_{cc}$ state from the  $DD^*$ molecular perspective,  together with SU(3)-flavor symmetry partners $D^{*0}D_s^{+}/D^{0}D_s^{*+}$ and $D^{*+}D_s^{+}/D^{+}D_s^{*+}$, as well as their HQSS partners $D^*D^*$, $D^{*0}D_s^{*+}/D^{*0}D_s^{*+}$,  and $D^{*+}D_s^{*+}/D^{*+}D_s^{*+}$. The multiplet hadronic molecule picture provides a nice approach to classifying exotic states. A notable example is the  three pentaquark states $P_{c}(4312)$, $P_{c}(4440)$, and $P_{c}(4457)$, which are well arranged into  the $\bar{D}^{(*)}\Sigma_c^{(*)}$ HQSS multiplet molecular picture~\cite{LHCb:2019kea,Liu:2019tjn}. Moreover, the multiplet hadronic molecule picture is argued that    experimental searches for their symmetry partners are helpful to verify the molecular nature of the exotic states. A notable example is that assuming $X(3872)$ as a $J^{PC}=1^{++}$ $\bar{D}^*D$ molecule, 
the experimental search for its HQSS partner $J^{PC}=2^{++}$ $\bar{D}^*D^*$ and  $J^{PC}=1^{++}$ $\bar{B}^*B$ are another model independent approach to verifying its molecular nature~\cite{Nieves:2012tt,Guo:2013sya}.  Along the same line, the molecular picture of $T_{cc}$ could also be tested by investigating its symmetry partners~\cite{Liu:2024uxn}.  To guide future experimental observations, we further investigate the partial decay properties of the $T_{cc}$ state and its symmetry partners, along with their production mechanisms in $B_c$ decays~\footnote{Different from the production of hidden-charm states in $b$-flavored hadron decays, doubly charmed tetraquarks can  be produced through $B_c$ decays.}. By considering the branching fractions of their partial decays and their production rates in $B_c$ decays, we identify the most promising channels for searching for doubly charmed tetraquark states. At last,  using the detection efficiency and the $B_{c}$ cross sections at the LHC, we estimate the event number of observing the doubly charmed tetraquark states.  

This work is structured as follows. In Section II, we begin with a brief introduction to the contact-range potentials for the doubly charmed tetraquark system. We then present the tree-level Feynman diagrams responsible for the three-body partial decays of doubly charmed hadronic molecules, as well as the triangle diagrams describing the two-body partial decays of doubly charmed tetraquark molecules and their production in $B_c$ meson decays. The effective Lagrangian approach is also outlined in this section. Numerical results and corresponding discussions are provided in Section III. Finally, a summary is presented in the last section.

\section{Theoretical formalism}

In this work, we focus on the doubly heavy charmed tetraquark with the quark configuration $cc \bar{q}\bar{q}$, where the light quark constituents are constrained by SU(3)-flavor symmetry. Since a charmed meson can be denoted by the $\bar{3}$ representation of the SU(3) flavor symmetry, a pair of charmed mesons contains irreducible representations ${3}$ and $\bar{6}$~\cite{Kaeding:1995vq}, where the $3$ representation includes three states with the flavor wave function  $I=0$~$DD^*$,  $I=1/2$~$\frac{1}{\sqrt{2}}(D_sD^*-D_s^*D)$, and the $\bar{6}$ representation includes six states with the flavor wave function $I=1$~$DD^*$,  $I=1/2$~$\frac{1}{\sqrt{2}}(D_sD^*+D_s^*D)$, and $I=0$~$D_sD_s^*$.  From the product of two matrix elements, the potential of the $3$ multiplet is attractive, but the potential of the $\bar{6}$ multiplet is repulsive~\cite{Chen:2024bre}.  The doubly charmed tetraquark state $T_{cc}$ can be assigned as a bound state of $I=0$~$DD^*$, and therefore   we  focus on the three states in the $3$ representation:  $I=0$~$D^{+}D^{*0}$,  $I=1/2$~$\frac{1}{\sqrt{2}}(D_s^+D^{*0}-D_s^{*+}D^0)$,  and $I=1/2$~$\frac{1}{\sqrt{2}}(D_s^+D^{*+}-D_s^{*+}D^+)$(denoted by $T_{cc}$, $T_{ccs}^{+}$, and $T_{ccs}^{++}$),  as well as their HQSS partners $I=0$~$D^{*+}D^{*0}$,  $I=1/2$~$\frac{1}{\sqrt{2}}(D_s^{*+}D^{*0}-D_s^{*+}D^{*0})$,  and $I=1/2$~$\frac{1}{\sqrt{2}}(D_s^{*+}D^{*+}-D_s^{*+}D^{*+})$(denoted by $T_{cc}^*$, $T_{ccs}^{*+}$, and $T_{ccs}^{*++}$).   
Taking into account the HQSS and SU(3)-flavor symmetry, the EFT  contact  potentials between a pair of heavy mesons can be parameterized  as~\cite{Liu:2019stu,Du:2021zzh,Peng:2023lfw} 
\begin{eqnarray}
    \label{potential}
 V(I=0, \, D_{(s)}D^*)&=& C_a +C_b,   \\ \nonumber
  V(I=0, \, D_{(s)}^*D^*)&=& C_a +C_b,
\end{eqnarray}
where $C_a$ and $C_b$  are unknown low-energy constants.  One can see that the contact range potentials between those six states are the same. Therefore, assuming the doubly charmed tetraquark state $T_{cc}$ as a $DD^*$ bound state, we can determine the contact potentials of  $I=0$~$DD^*$, and then predict the masses of the other five states. In this work, we employ the contact range EFT approach to study the spectrum of these doubly charmed tetraquark states~\cite{Liu:2018zzu,Wu:2025fzx}. As shown later, in addition to $T_{cc}$,  we predict five more doubly charmed tetraquark states. In the following, we investigate the decays and production of these tetraquark states in $B_c$ meson decays.   

\begin{figure}[htbp]
  \centering
  \includegraphics[width=0.55\textwidth]{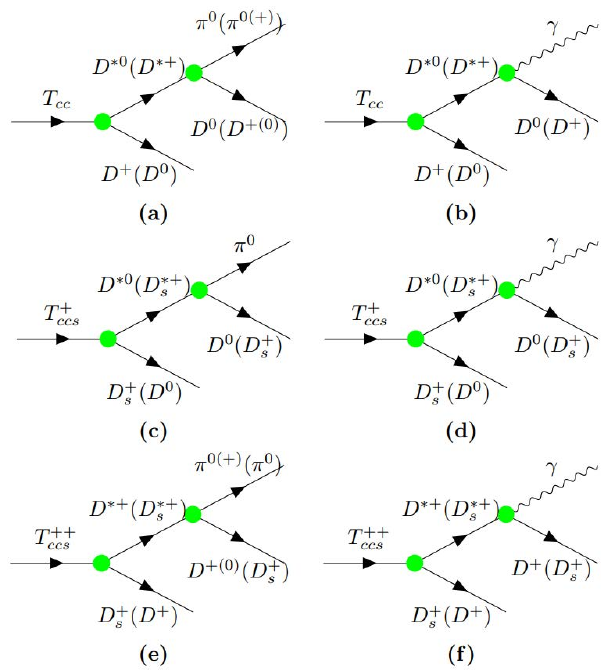}
\caption{Tree diagrams responsible for the three-body decays of the $T_{cc}$(a-b), $T_{ccs}^{+}$(c-d), and $T_{ccs}^{++}$(e-f) states.   }
\label{treeDDs}
\end{figure}

\begin{figure}[htbp]
  \centering
  \includegraphics[width=0.55\textwidth]{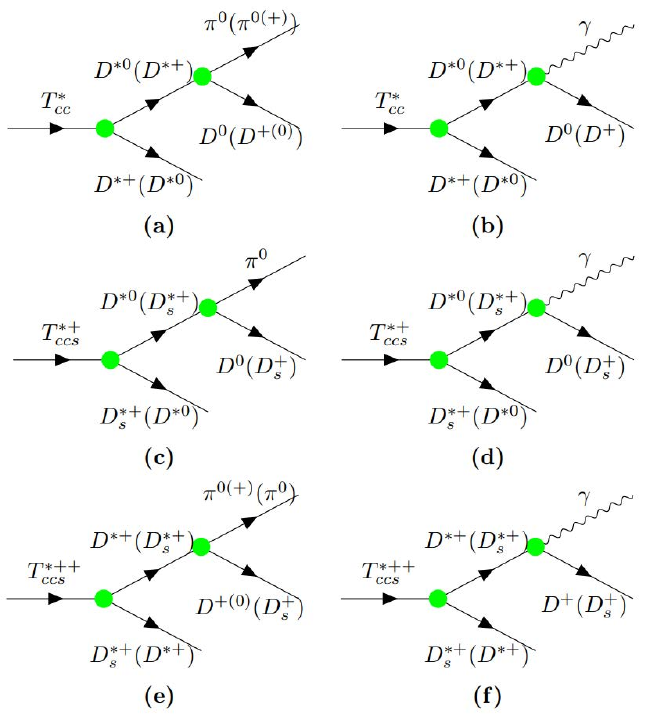}
\caption{Tree diagrams responsible for the three-body decays of the $T_{cc}^*$(a-b), $T_{ccs}^{*+}$(c-d), and $T_{ccs}^{*++}$(e-f) states.   }
\label{DsDs}
\end{figure}

\begin{figure}[htbp]
  \centering
  \includegraphics[width=0.7\textwidth]{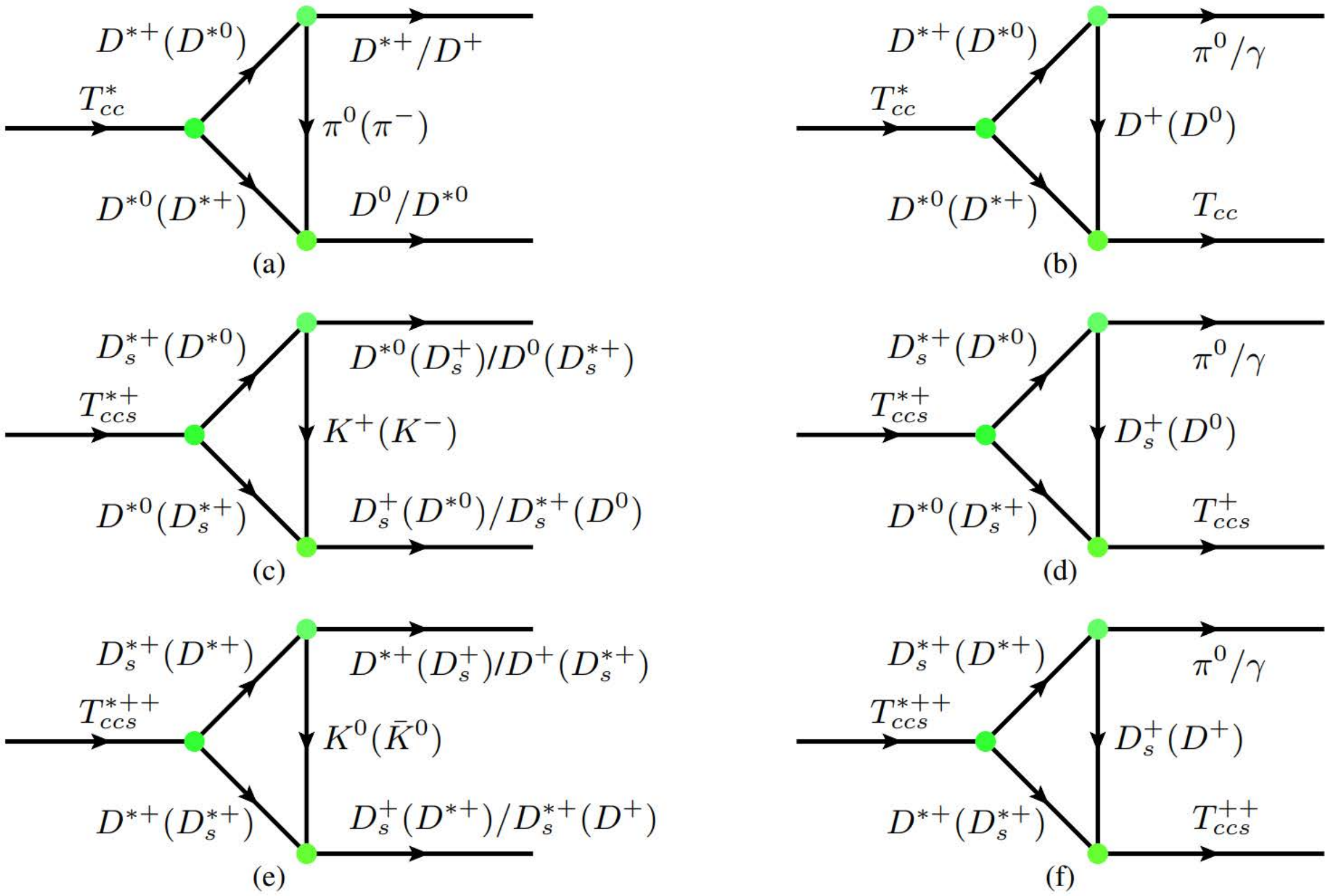}
\caption{Triangle diagrams responsible for the two-body decays of the $T_{cc}^*$(a-b), $T_{ccs}^{*+}$(c-d), and $T_{ccs}^{*++}$(e-f) states.   }
\label{triangle1}
\end{figure}

For the $T_{cc}$, $T_{ccs}^+$, and $T_{ccs}^{++}$ states, two-body strong decays are  forbidden. Instead, they can decay via the strong or radiative decays of the vector charmed mesons, proceeding via the tree diagrams as illustrated in Fig.~\ref{treeDDs}. Since only the three-body decays are allowed, the widths of the $T_{cc}$, $T_{ccs}^+$, and $T_{ccs}^{++}$ states are expected to be less than the widths of the $D_{(s)}^*$ mesons. Similarly, the  $T_{cc}^*$, $T_{ccs}^{*+}$, and $T_{ccs}^{*++}$ states can proceed via the three-body decay modes as shown in Fig.~\ref{DsDs}. Moreover, the $T_{cc}^*$, $T_{ccs}^{*+}$, and $T_{ccs}^{*++}$ states can undergo two-body decays through triangle diagrams, as illustrated in Fig.~\ref{triangle1}. In addition to the $DD_s^{*}$ channel, these states may also decay into their HQSS partners, accompanied by a pion or a photon. These HQSS partners subsequently decay into three-body final states, as shown in Fig.~\ref{treeDDs}, which effectively correspond to four-body decays of the $T_{cc}^*$, $T_{ccs}^{*+}$, and $T_{ccs}^{*++}$ states.   Since the two-body decays are allowed, the widths of the  $T_{cc}^*$, $T_{ccs}^{*+}$, and $T_{ccs}^{*++}$ states are expected to be larger than their HQSS partners.

\begin{figure}[htbp]
  \centering
  \includegraphics[width=1.0\textwidth]{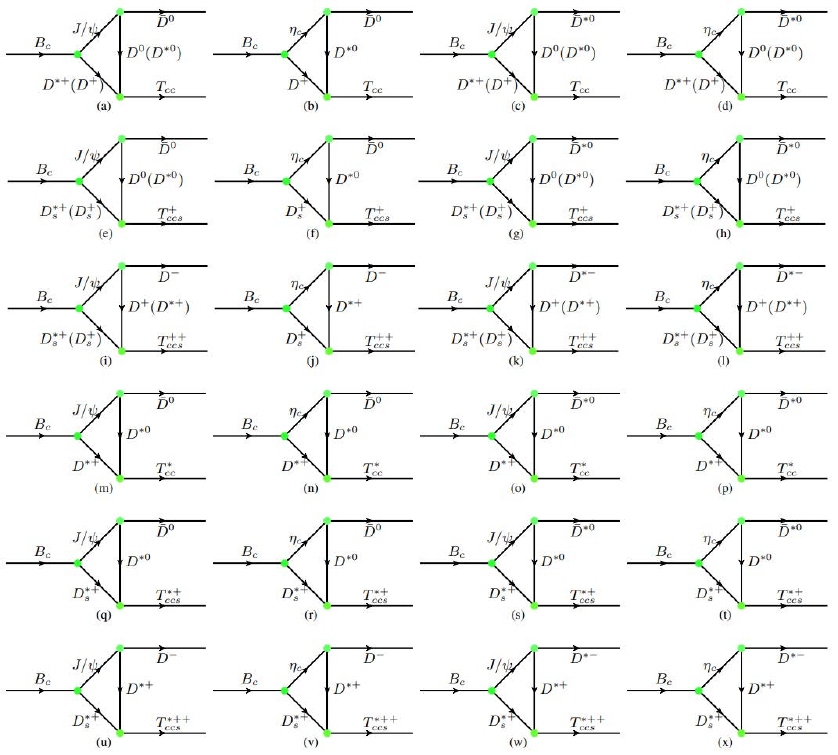}
\caption{Triangle diagrams responsible for the productions of the $T_{cc}$(a-d), $T_{ccs}^+$(e-h), $T_{ccs}^{++}$(i-l), $T_{cc}^*$(m-p), $T_{ccs}^{*+}$(q-t), and  $T_{ccs}^{*++}$(u-x) states in $B_c$ meson decays.   }
\label{triangle2}
\end{figure}

Unlike the production of $T_{cc}$ in inclusive processes~\cite{LHCb:2021vvq,LHCb:2021auc},  we study the production of the doubly charmed tetraquark states in $B_c$ meson decays in this work. 
Heavy-flavor hadron decays are a primary way to produce exotic hadron states. For instance, many hidden-charm exotic states were produced in b-flavored hadron decays. The production of these doubly charmed tetraquark states in $B_c$ meson decays is difficult to calculate at the quark level, but can be calculated at the hadron level via the FSIs approach.   Recent studies indicated that FSIs play an important role in producing exotic states~\cite{Wu:2019rog,Wu:2023rrp,Yu:2023avh,Yuan:2025pnt}. 
As shown in Fig.~\ref{triangle2}, the $B_c$ meson first weakly decays into charmonium states($J/\psi$ and $\eta_c$) and charmed mesons $D_{(s)}^{(*)}$, then the charmonium states scatter into a pair of charmed mesons $D^{(*)}\bar{D}^{(*)}$, and finally the doubly charmed tetraquark states are dynamically generated via the $D^{(*)}D_{(s)}^{(*)}$ interactions. In this work, we neglect the contributions of orbitally and radially excited charmonium states, which are relatively small due to the suppression of $B_{c}$ meson transition to excited charmonium states and the couplings of the excited charmonium states to a pair of charmed mesons.

\subsection{Effective Lagrangians }

This section presents the effective Lagrangians that will be utilized in this work for computing the Feynman diagrams depicted in Figs.~\ref{treeDDs}-\ref{triangle2}.   Since the Lagrangians involving $D^{(*)}$ mesons are similar to those with $D_s^{(*)}$ mesons via SU(3)-flavor symmetry, we only present the Lagrangians for the charmed mesons $D^{(*)}$ here.   The first category of Lagrangians involves the interactions between the hadronic molecules and their constituents.   
The interactions of the $T_{cc}$ state  with $DD^*$ and the $T_{cc}^*$ state  with $D^*D^*$ are described by the following Lagrangian~\cite{Wu:2023rrp}:
\begin{eqnarray}
\mathcal{L}_{T_{cc}D^*D}&=& g_{T_{cc}D^*D} T_{cc}^{\mu}  D  D^{\ast}_{\mu}, 
 \\ \nonumber
\mathcal{L}_{T_{cc}^*D^*{D}^*}&=& i g_{T_{cc}^*D^*{D}^*}\varepsilon_{\mu\nu\alpha\beta} \partial^{\mu}T_{cc}^{*\nu}D^{\ast\alpha} {D}^{\ast\beta},
\end{eqnarray}
where   $g$  with a subscript denotes the couplings between the hadronic molecules and their corresponding constituents. These couplings are determined by the contact range approach~\cite{Wu:2025fzx}. After obtaining the pole positions of these doubly charmed tetraquark states, we determined the couplings as   $g_{T_{cc}D^{*+}D^0}=4.12$~GeV, $g_{T_{cc}D^{*0}D^+}=3.67$~GeV, $g_{T_{ccs}^+D^{*0}D_s^+}=4.87$~GeV, $g_{T_{ccs}^+D^{0}D_s^{*+}}=4.28$~GeV,  $g_{T_{ccs}^{++}D^{*+}D_s^+}=4.51$~GeV, $g_{T_{ccs}^{++}D^{+}D_s^{*+}}=3.66$~GeV, $g_{T_{cc}^*D^*D^*}=1.92$,  $g_{T_{ccs}^{*+}D_s^{*+}D^{*0}}=2.09$, and $g_{T_{ccs}^{*++}D_s^{*+}D^{*+}}=2.08$.

The Lagrangian describing  the interactions between  charmed mesons and $\pi/\gamma$ is written as~\cite{Oh:2000qr}
\begin{eqnarray}
\mathcal{L}_{D D^{\ast} \pi}&=& -i g_{D D^{\ast} \pi} (D \partial^{\mu}\pi D^{\ast\dag}_{\mu}-D_{\mu}^* \partial^{\mu} \pi  D^{\dag}),  \\ \nonumber
\mathcal{L}_{D^{\ast} D^{\ast} \pi}&=& - g_{D^{\ast} D^{\ast} \pi} \varepsilon_{\mu\nu\alpha\beta} \partial^{\mu}D^{\ast\nu} {\partial}^{\alpha} {D}^{\ast\beta\dag} \pi ,   \\ \nonumber 
\mathcal{L}_{D D^{\ast} \gamma}&=&e g_{D D^{\ast} \gamma}\varepsilon^{\mu\nu\alpha\beta} \partial_{\mu}A_{\nu}\partial_{\alpha}D^{\ast\dag}_{\beta}D,    
\end{eqnarray}
where the  relevant couplings are taken  as $g_{ D^{\ast+}D^{0}\pi^{+} }=16.818$ and  $g_{ D^{\ast+}D^{+}\pi^{0} }=11.940$,  $g_{D^{\ast+} D^{+}  \gamma}=$0.468 GeV$^{-1}$, and $g_{D^{\ast0} D^{0}  \gamma}=$1.843 GeV$^{-1}$~\cite{Ling:2021bir}. The  electron charge   $e$ is  derived by the relationship  $\frac{e^2}{4\pi}=\frac{1}{137}$.      The couplings $g_{D^*D^*\pi}$ are determined by the relationship  $g_{D^*D^*\pi}=g_{D^*D\pi}/m_{\bar{D}}$, where $m_{\bar{D}}$ is the average mass of $D$ and $D^*$~\cite{Oh:2000qr,Irfan:2015qxa}. The couplings of their SU(3)-flavor partners are determined as $g_{ D^{\ast}D_sK }=15.275$ and $g_{ D^{\ast}D_s^*K }=7.887$~GeV$^{-1}$~\cite{Colangelo:2003sa}.     
If the partial decay width of $D_s^{*+} \to D_s^+ \gamma$ is chosen to be  $0.066$ keV from the lattice QCD calculation~\cite{Donald:2013sra}, we can determine the coupling  $g_{ D_s^{\ast+}D_s^{+}\gamma }=0.101$~GeV$^{-1}$.  According to the ratio of the partial decay width $\Gamma(D_s^{\ast+} \to D_s^{+}\gamma)$ to that of $\Gamma(D_s^{\ast+} \to D_s^{+}\pi^0)$~\cite{ParticleDataGroup:2024cfk},   we  can determine the coupling $g_{ D_s^{\ast+}D_s^{+}\pi^{0} }=0.111$.

The effective  Lagrangians describing the interactions between charmonium states ($J/\psi$, $\eta_{c}$) and a pair of charmed mesons read~\cite{Oh:2000qr,Colangelo:2003sa,Guo:2014taa,Shi:2023mer}  
\begin{eqnarray}
\mathcal{L}_{\psi DD}&=&ig_{\psi DD} \psi_{\mu}(\partial^{\mu}D\bar{D}-D\partial^{\mu}\bar{D}), \\ \nonumber 
\mathcal{L}_{\psi DD^{\ast}}&=&-g_{\psi DD^{\ast}} \epsilon^{\alpha \beta \mu \nu}\partial_{\alpha} \psi_{\beta}(\partial_{\mu} D_{\nu}^{\ast}\bar{D}+ D \partial_{\mu}\bar{D}^{\ast}_{\nu}), 
\\ \nonumber 
\mathcal{L}_{\psi D^{\ast}D^{\ast}}&=&-ig_{\psi D^{\ast}D^{\ast}} [\psi^{\mu}(\partial_{\mu}D^{\ast}_{\nu}\bar{D}^{\ast\nu}-D^{\ast \nu}\partial_{\mu}\bar{D}^{\ast}_{\nu}) +(\partial_{\mu}\psi_{\nu}D^{\ast\nu}-\psi_{\nu}\partial_{\mu}D^{\ast\nu})\bar{D}^{\ast\mu} \\  \nonumber
&&+ {D}^{\ast\mu} (\psi^{\nu}\partial_{\mu}\bar{D}^{\ast}_{\nu}-\partial_{\mu}\psi_{\nu}\bar{D}^{\ast\nu})],  \\ \nonumber
\mathcal{L}_{\eta_{c} D^{\ast}D}&=&ig_{\eta_{c} D^{\ast}D}[D^{\ast\mu}(\partial_{\mu}\eta_{c}\bar{D}-\eta_{c}\partial_{\mu}\bar{D})-(\partial_{\mu}\eta_{c}D-\eta_{c}\partial_{\mu}D)\bar{D}^{\ast\mu}],   \\ \nonumber
\mathcal{L}_{\eta_{c} D^{\ast}D^{\ast}}&=&-g_{\eta_{c} D^{\ast}D^{\ast}}\varepsilon^{\mu\nu\alpha\beta}\partial_{\mu}D^{\ast}_{\nu}{\partial}_{\alpha} \bar{D}^{\ast}_{\beta}\eta_{c}, 
\end{eqnarray}
where $g_{\psi DD}$,  $g_{\psi D^{\ast}D}$,  $g_{\psi D^{\ast}D^\ast}$, $g_{\eta_{c} D^{\ast}D}$, and $g_{\eta_{c} D^{\ast}D^{\ast}}$ are the couplings of the charmonium mesons to the charmed mesons. The coupling constants   are determined as follows:  $g_{\psi DD}=g_{\psi D^*D^*}=m_{\psi}/f_{\psi}$, $g_{\psi D^*D}=\frac{1}{m_{\bar{D}}}g_{\psi DD}$~\cite{Deandrea:2003pv,Oh:2007ej}, $g_{\eta_{c} D^* D}=g_{\eta_{c} D^*D^*}\sqrt{\frac{m_D}{m_{D^*}}}m_{\eta_c}=g_{2}\sqrt{m_{\eta_c}m_{D}m_{D^*}} $ with $g_{2}=\frac{\sqrt{m_{\psi}}}{2m_D f_{\psi}}$~\cite{Wang:2012wj}.

The weak decays $B_c \to \eta_c D_{(s)}^{(*)}$ and $B_c \to J/\psi D_{(s)}^{(*)}$ proceed mainly via the external $W$-emission mechanism at the quark level, which belongs to $T-$type diagrams and is the dominant contribution in terms of the topological classification of weak decays~\cite{Chau:1987tk,Ali:1998eb,Ali:2007ff,Li:2012cfa}.
In the naive factorization ansatz~\cite{Bauer:1986bm}, the amplitudes of the weak decays $B_c \to \eta_c D_{(s)}^{(*)}$ and $B_c \to J/\psi D_{(s)}^{(*)}$ can be expressed as products of two current matrix elements
\begin{eqnarray}\label{Bs-M}
\mathcal{A}\left(B_c \to \eta_c D_{(s)} \right)&=&\frac{G_{F}}{\sqrt{2}} V_{cb}V_{cd(s)} a_{1}\left\langle D_{(s)}|J_{\mu}| 0\right\rangle\left\langle \eta_c|J^{\mu}| B_c\right\rangle, \\
\mathcal{A}\left(B_c \to \eta_c D_{(s)}^* \right)&=&\frac{G_{F}}{\sqrt{2}} V_{cb}V_{cd(s)} a_{1}\left\langle D_{(s)}^*|J_{\mu}| 0\right\rangle\left\langle \eta_c|J^{\mu}| B_c\right\rangle, \\
\mathcal{A}\left(B_c \to J/\psi D_{(s)} \right)&=&\frac{G_{F}}{\sqrt{2}} V_{cb}V_{cd(s)} a_{1}\left\langle D_{(s)}|J_{\mu}| 0\right\rangle\left\langle J/\psi|J^{\mu}| B_c\right\rangle, \\
\mathcal{A}\left(B_c \to J/\psi D_{(s)}^* \right)&=&\frac{G_{F}}{\sqrt{2}} V_{cb}V_{cd(s)} a_{1}\left\langle D_{(s)}^*|J_{\mu}| 0\right\rangle\left\langle J/\psi|J^{\mu}| B_c\right\rangle,
\end{eqnarray}
where $J_\mu=\bar{q}_1 \gamma_\mu(1-\gamma_5)q_2$,  $G_F$ is the Fermi coupling constant, $V_{cb}$ and $V_{cd(s)}$ are the Cabibbo-Kobayashi-Maskawa (CKM) matrix elements, and $a_{1}$ is the Wilson coefficient depending on the renormalization scale~\cite{Chau:1987tk,Cheng:1993gf,Cheng:2010ry}. In this work, we deal with $B_c$ decays, and therefore extract the value of the effective Wilson coefficient at the energy scale $\mu=m_b$, i.e., $a_1=1.07$~\cite{Li:2012cfa}, consistent with Ref.~\cite{Lu:2009cm}. Moreover, the weak decays $B_c \to \eta_c D_{(s)}^{(*)}$ and $B_c \to J/\psi D_{(s)}^{(*)}$ proceed via the  internal $W-$emission mechanism at the quark level,  which belong to $C$-type diagrams.  The amplitudes of the weak decays $B_c \to \eta_c D_{(s)}^{(*)}$ and $B_c \to J/\psi D_{(s)}^{(*)}$ belonging to $C$-type diagrams are color suppressed and neglected, and therefore only those of $T$-type diagrams are taken into account.

The former current matrix elements are written as
\begin{eqnarray}
\langle0|\bar{q}_1 \gamma_\mu \gamma_5 q_2|D_{(s)}(p_1)\rangle&=&-i f_{D_{(s)}} p_{1\mu},  \nonumber \\
\langle0|\bar{q}_1 \gamma_\mu q_2|D_{(s)}^*(p_1,\epsilon)\rangle&=&f_{D_{(s)}^*} \epsilon_\mu(p_1) m_{D_{(s)}^*},
\label{Eq:ME1}
\end{eqnarray}
where $f_{D_{(s)}}$ and $f_{D_{(s)}^*}$ are the decay constants of $D_{(s)}$ and $D_{(s)}^*$. In this work, we take  $G_F = 1.166 \times 10^{-5}~{\rm GeV}^{-2}$, $V_{cb}=0.041$,   $V_{cs}=0.973$, $V_{cd}=0.226$, $a_1=1.07$, $f_D = 210.4$ MeV,  $f_{D^{\ast}}=250$~MeV,  $f_{D_s} = 247.7$ MeV, and $f_{D_s^{\ast}}=282.5$~MeV~ \cite{ParticleDataGroup:2022pth,Verma:2011yw,FlavourLatticeAveragingGroupFLAG:2024oxs,Narison:2015nxh,Li:2017mlw,Li:2012cfa,Wang:2015mxa}.

The latter current matrix elements describing the transitions of $B_c\to \eta_c$ and $B_c\to J/\psi$ are characterized by a series of form factors~\cite{Verma:2011yw,FlavourLatticeAveragingGroupFLAG:2024oxs}:
\begin{eqnarray}\label{eq:etaC_ff}
\langle \eta_c(p_{\eta_c})| \bar{c} \gamma_\mu b | B_c(p_{B_c}) \rangle &=& \left [ P_\mu - \frac{m_{B_c}^2 - m_{\eta_c}^2}{q^2} q_\mu \right]f_+(q^2) + \left [ \frac{m_{B_c}^2 - m_{\eta_c}^2}{q^2} q_\mu\right ] f_0 (q^2)\,,
\\
\langle J/\psi(p_{J\psi},\epsilon) | \bar c\gamma_\mu(1-\gamma_5) b | B_c(p_{B_c})\rangle & =&
i\epsilon_{\alpha}^{*}\{-g^{\mu \alpha} (m_{B_c}+m_{J/\psi}) A_{1}\left(q^{2}\right)+P^{\mu} P^{\alpha} \frac{A_{2}\left(q^{2}\right)}{m_{B_c}+m_{J/\psi}} \\  \nonumber
&&+i \varepsilon^{\mu \alpha \beta \gamma}P_\beta q_\gamma \frac{V\left(q^{2}\right)}{m_{B_c}+m_{J/\psi}} \\ \nonumber && +q^{\mu} P^{\alpha} \left[\frac{m_{B_c}+m_{J/\psi}}{q^{2}}A_{1}\left(q^{2}\right)-\frac{m_{B_c}-m_{J/\psi}}{q^{2}}A_{2}\left(q^{2}\right)-\frac{2m_{J/\psi}}{q^{2}}A_{0}\left(q^{2}\right)\right]\}, 
\label{eq:FF1}
\end{eqnarray}
where $\epsilon$ is the polarization vector of the $J/\psi$ meson, $q=p_{B_c}-p_{\eta_c(J/\psi)}$ is the momentum transfer, and $V(q^2)$, $A_0(q^2)$, $A_1(q^2)$ and $A_2(q^2)$ are the transition form factors. In general, the transition form factors extracted from the light cone sum rules can be extrapolated by using Bourrely-Caprini-Lellouch (BCL) parametrization \cite{Bourrely:2008za} of the form factor series expansion in powers of a conformal mapping variable, which satisfies unitarity, analyticity, and perturbative QCD scaling. The BCL parametrization is based on a rapidly converging series in the parameter $z$ as
\begin{eqnarray}
f(t) &=& \frac{1}{P(t)} \sum_{k=0} \alpha_k [z(t)-z(0)]^k, \nonumber \\
z( t)  &=&  \frac{ \sqrt{t_+ - t} - \sqrt{t_+ - t_0}}{\sqrt{t_+ - t} + \sqrt{t_+ - t_0}} ,
\label{z-exp}
\end{eqnarray}
weighted by a simple pole function $P(t) = 1- t/m_R^2$ that accounts for low-lying resonances present below the threshold production of real $B_c-M$ pairs at $t_+ = (m_{B_c} + m_M)^2$.
The parameter $t_0$, $0 \leq t_0 \leq t_- = (m_{B_c} - m_M)^2$ is a free parameter that can be used to optimize the convergence of the series expansion.  For the truncation to only two terms in expansion Eq.~(\ref{z-exp}), it was shown that the optimized value of $t_0$ has the form~\cite{Bharucha:2010im}:
\begin{eqnarray}
t_{0{|\rm opt}} =  t_+  \left (1 - \sqrt{1 - \frac{t_-}{t_+}} \right ).
\end{eqnarray}
and that the other choices of $t_0$ do not make a visible change in the form factor parameterization. The masses of resonances appearing in the fits are determined by the properties of the form factors.  In Table~\ref{tab:BCLfit}, we collect the values of the fitting parameters in the form factors to be used in this work.

\begin{table*}[ttt]
\centering
\caption{Summary of the BCL fit for $B_c \to \eta_c$  and $B_c \to J/\psi$ transition form factors. \label{tab:BCLfit}}
 \begin{tabular}{ c | c  c  c  c }
 \hline\hline
~~~~~Form factor~~~~~ &  ~~~~~~~~$J^P$~~~~ & ~~~$m_R$  (GeV)~~~ & ~~~~~$\alpha_0$~~~~~ & ~~~~~~~~$\alpha_1$~~~~~ \\
 \hline
 $f_+$ & $1^-$& 6.34 & 0.62 & -6.13 \\
$f_0$ & $0^+$ & 6.71 & 0.63 & -4.86 \\
\hline
$V$ & $1^-$ & 6.34 & 0.74 & -8.66 \\
$A_0$ & $0^-$ & 6.28 & 0.54 & -6.80 \\
$A_1$ & $1^+$ & 6.75 & 0.55 & -4.67 \\
$A_2$ & $1^+$ & 6.75 & 0.35 & -1.78 \\
\hline\hline
\end{tabular}
\end{table*}

With Eqs.(\ref{Eq:ME1})-(\ref{eq:FF1}), we obtain the amplitudes for the decays $B_c(k_0) \to D^{(*)}(q_1) \eta_c/J/\psi(q_2)$,

\begin{eqnarray}\label{Bs-M}
\mathcal{A}\left(B_c \to \eta_c D \right)&=& i \frac{G_{F}}{\sqrt{2}} V_{cb}V_{cd(s)} a_{1} f_D (m_{B_c}^2-m_{\eta_c}^2) f_0(q^2), \\
\mathcal{A}\left(B_c \to \eta_c D^* \right)&=&\frac{G_{F}}{\sqrt{2}} V_{cb}V_{cd(s)} a_{1} m_{D^\ast} f_{D^\ast} (k_0 + q_2) \cdot \epsilon(q_1) f_+(q^2), 
\end{eqnarray}
\begin{eqnarray}
\mathcal{A}\left(B_c \to J/\psi D \right)&=&- \frac{G_{F}}{\sqrt{2}} V_{cb}V_{cd(s)} a_{1} f_D \bigg\{-q_1 \cdot \epsilon(q_2) (m_B+m_{J/\psi}) A_1(q^2) +(k_0+q_2) \cdot \epsilon(q_2) q_1 \cdot (k_0+q_2) \frac{A_2(q^2)}{m_B+m_{J/\psi}} \nonumber\\
&&{} +(k_0+q_2) \cdot \epsilon(q_2) \bigg[(m_B+m_{J/\psi})A_1(q^2) -(m_B-m_{J/\psi})A_2(q^2) -2 m_{J/\psi} A_0(q^2)\bigg]\bigg\}, \\
\mathcal{A}\left(B_c \to J/\psi D^* \right)&=&i \frac{G_{F}}{\sqrt{2}} V_{cb}V_{cd(s)} a_{1} m_{D^\ast} f_{D^\ast} \epsilon_\mu(q_1) \bigg[- g_{\mu\alpha} (m_B+m_{J/\psi})A_1(q^2) + (k_0+q_2)_\mu (k_0+q_2)_\alpha \frac{A_2(q^2)}{m_B+m_{J/\psi}} \nonumber\\
&&{} +i \epsilon_{\mu\alpha\rho\sigma} (k_0+q_2)^\rho q^\sigma \frac{V(q^2)}{m_B+m_{J/\psi}}\bigg] \epsilon^\alpha(q_2),
\end{eqnarray}

\subsection{  Three-body decay amplitudes  }
 
With the above Lagrangians, the decay amplitudes of $T_{cc}\to DD\pi$ and $T_{cc} \to DD\gamma$ of Fig.~\ref{treeDDs} are written as 
\begin{eqnarray}
\mathcal{M}_{T_{cc}\to DD\pi}^{a}&=& ig_{T_{cc}}g_{D D^{\ast} \pi}~p_{2\mu} \frac{-g^{\mu\nu}+k_{1}^{\mu}k_{1}^{\nu}/m_{D^{\ast}}^{2}}{k_{1}^{2}-m_{D^{\ast}}^{2}+im_{D^{\ast}}\Gamma_{m_{D^{\ast}}}}\varepsilon_{\nu}(p_{0}),  \\ 
\mathcal{M}_{T_{cc}\to DD\gamma}^{b}&=& ig_{T_{cc}}g_{D D^{\ast} \gamma}\varepsilon^{\mu\nu\alpha\beta} ~p_{2\mu}\varepsilon_{\nu}(p_{2})k_{1\alpha} \frac{-g_{\beta\sigma}+k_{1\beta}k_{1\sigma}/m_{D^{\ast}}^{2}}{k_{1}^{2}-m_{D^{\ast}}^{2}+im_{D^{\ast}}\Gamma_{m_{D^{\ast}}}}\varepsilon^{\sigma}(p_{0}), 
\end{eqnarray}
where $p_{2}$, $k_{1}$, and $p_{0}$ are the momentum of $\pi$($\gamma$), $D^{\ast}$, and $T_{cc}$, respectively.  Similarly, we express the decay amplitudes  of   $ T_{cc}^* \to DD^* \pi/\gamma $ and  $ T_{ccs}^* \to D_s^*D\pi/\gamma $  of  Fig.~\ref{DsDs}  
\begin{eqnarray}
\mathcal{M}_{T_{cc}^*\to D^*D\pi}^{a}&=& ig_{T_{cc}^*D^*D^*}g_{D D^{\ast} \pi}~\varepsilon_{\mu\nu\alpha\beta}p_{0}^{\mu}p_{2}^{\lambda} \frac{-g^{\alpha\lambda}+k_{1}^{\alpha}k_{1}^{\lambda}/m_{D^{\ast}}^{2}}{k_{1}^{2}-m_{D^{\ast}}^{2}+im_{D^{\ast}}\Gamma_{m_{D^{\ast}}}}\varepsilon_{\nu}(p_{0})\varepsilon_{\beta}(p_1),  \\ 
\mathcal{M}_{T_{cc}^*\to D^*D\gamma}^{b}&=& ig_{T_{cc}^*D^*D^*}g_{D D^{\ast} \gamma}\varepsilon^{\mu\nu\alpha\beta} ~p_{2\mu}\varepsilon_{\nu}(p_{2})k_{1\alpha} \frac{-g_{\beta b}+k_{1\beta}k_{1 b}/m_{D^{\ast}}^{2}}{k_{1}^{2}-m_{D^{\ast}}^{2}+im_{D^{\ast}}\Gamma_{m_{D^{\ast}}}}\varepsilon_{uvab}k_0^{u}\varepsilon^{v}(k_0)\varepsilon^{a}(p_1), 
\end{eqnarray}
where $p_1$, $p_{2}$, $k_{1}$, $p_3$ and $p_{0}$ are the momentum of $D$,  $\pi$($\gamma$), $D^{\ast}$, $D^{\ast}$, and $T_{cc}$, respectively. Here we note that the amplitudes of Fig.~\ref{treeDDs}~(a) and Fig.~\ref{DsDs}~(a) are similar to those of  Fig.~\ref{treeDDs}~(c,e) and Fig.~\ref{DsDs}~(c,e), and  Fig.~\ref{treeDDs}~(b) and Fig.~\ref{DsDs}~(b) are similar to those of  Fig.~\ref{treeDDs}~(d,f) and Fig.~\ref{DsDs}~(d,f).

As for the three-body decays, the partial decay widths of   $T_{cc(s)}^{(*)} \to D_{(s)} D_{(s)}^{(*)} \pi/\gamma$    as a function of $m_{12}^2$ and $m_{23}^2$ read
\begin{equation}
\Gamma =  \frac{1}{(2 \pi)^{3}}\frac{1}{2J+1} \int\int \frac{\overline{|\mathcal{M}|^2}}{32 m_{T_{cc(s)}^{(*)}}^{3}} d m_{12}^{2} d m_{23}^{2},
\end{equation}
with $m_{12}$ the invariant mass of $D_{(s)} D_{(s)}^{(*)}$  and $m_{23}$ the invariant mass of $D_{(s)}^{(*)}\pi$ or $D_{(s)}^{(*)}\gamma$ for the $T_{cc(s)}^{(*)} \to D_{(s)} D_{(s)}^{(*)} \pi/\gamma$   decays.

\subsection{ Two-body decay amplitudes}

The decay amplitudes  of   $ T_{cc}^* \to DD^* $ and  $ T_{cc}^* \to T_{cc}\pi/\gamma $  of Fig.~\ref{triangle1} are written as 
 \begin{eqnarray}
 \label{gamma}   
 i\mathcal{M}^{a}_{T_{cc}^* \to DD^* }&=&g_{T_{cc}^*D^*D^*} g_{DD^*\pi}g_{D^*D^*\pi} \int \frac{d^{4}q}{(2\pi)^4} \frac{-g^{\theta\alpha}+\frac{k_1^\theta k_1^\alpha}{k_1^2}}{k_1^2-m_{D^*}^2}  
\frac{1}{q^{2}-m_{\pi}^{2}}  \frac{-g^{\beta b}+\frac{k_2^\beta k_2^b}{k_2^2}}{k_2^{2}-m_{D^*}^{2}} \\ \nonumber   && [-\varepsilon_{\mu\nu\alpha\beta} k_0^{\mu}\varepsilon^{\nu}(k_0)][-q^{\theta}][-\varepsilon_{uvab}p_{2}^{u}\varepsilon^{v}(p_2)k_2^a]F^2(q^2),   \\ \nonumber 
 i\mathcal{M}^{b}_{T_{cc}^* \to T_{cc} \pi^0 }&=&g_{T_{cc}^*D^*D^*} g_{DD^*\pi}g_{T_{cc}D^*D} \int \frac{d^{4}q}{(2\pi)^4} \frac{-g^{\theta\alpha}+\frac{k_1^\theta k_1^\alpha}{k_1^2}}{k_1^2-m_{D^*}^2}  
\frac{1}{q^{2}-m_{D}^{2}}  \frac{-g^{\beta a}+\frac{k_2^\beta k_2^a}{k_2^2}}{k_2^{2}-m_{D^*}^{2}} \\ \nonumber   && [-\varepsilon_{\mu\nu\alpha\beta} k_0^{\mu}\varepsilon^{\nu}(k_0)][-p_1^{\theta}][\varepsilon^{a}(p_2)]F^2(q^2),   \\ \nonumber 
 i\mathcal{M}^{b}_{T_{cc}^* \to T_{cc} \gamma}&=&g_{T_{cc}^*D^*D^*} g_{DD^*\gamma}g_{T_{cc}D^*D} \int \frac{d^{4}q}{(2\pi)^4} \frac{-g^{b\alpha}+\frac{k_1^b k_1^\alpha}{k_1^2}}{k_1^2-m_{D^*}^2}  
\frac{1}{q^{2}-m_{D}^{2}}  \frac{-g^{\beta \theta}+\frac{k_2^\beta k_2^\theta}{k_2^2}}{k_2^{2}-m_{D^*}^{2}} \\ \nonumber   && [-\varepsilon_{\mu\nu\alpha\beta} k_0^{\mu}\varepsilon^{\nu}(k_0)][e \varepsilon_{uvab}p_1^{u}\varepsilon^{v}(p_1)k_1^{a}][\varepsilon^{\theta}(p_2)]F^2(q^2),
 \end{eqnarray}
where $q$ and $p_1$ represent the momenta for the $D$ meson and the $T_{cc}$ state, and $\varepsilon_\mu$ represent the  polarization vector for the state  of spin $S=1$. Here one should note that  the amplitudes of Fig.~\ref{triangle1}~(a) and Fig.~\ref{triangle1}~(b) are similar to those of Fig.~\ref{triangle1}~(c,e) and Fig.~\ref{triangle1}~(d,f), respectively.

Similarly, the amplitudes of the decays $B_c \to T_{cc(s)} \bar{D}^{(*)}$ in Fig.~\ref{triangle2}  are written as
\begin{eqnarray}
\mathcal{M}_{a,e,i}&=&i^2 \int\frac{d^4 q}{(2\pi)^4}\Big[\mathcal{A}_{\mu\nu}(B_c\to J/\psi D_{(s)}^{*})\Big]\Big[g_{T_{cc}D^*D}\varepsilon_{\beta}(p_2)\Big]\Big[g_{ \psi \bar{D}D}(q-p_{1})_{\alpha}\Big]  \nonumber  \\  
&&\frac{-g^{\mu\alpha}+q^\mu_1 q^\alpha_1 /q^2_1}{q^2_1-m^2_1}\frac{-g^{\nu\beta}+q^\nu_2 q^\beta_2 /q^2_2}{q^2_2-m^2_2}\frac{1}{q^2-m^2}\mathcal{F}(q^2,m^2),  \\ \nonumber
\mathcal{M}_{a,e,i}&=&i^2 \int\frac{d^4 q}{(2\pi)^4}\Big[\mathcal{A}_{\mu}(B_c\to J/\psi D_{(s)})\Big]\Big[g_{T_{cc}D^*D}\varepsilon_{\beta}(p_2)\Big]\Big[-g_{ \psi \bar{D}D^*}\varepsilon^{\sigma\alpha\tau\nu}q_{1\sigma}q_{\tau}\Big]  \nonumber  \\  
&&\frac{-g^{\mu\alpha}+q^\mu_1 q^\alpha_1 /q^2_1}{q^2_1-m^2_1}\frac{1}{q^2_2-m^2_2}\frac{-g^{\nu\beta}+q^\nu q^\beta /q^2}{q^2-m^2}\mathcal{F}(q^2,m^2),  \\ \nonumber
\mathcal{M}_{b,f,j}&=&i^2 \int\frac{d^4 q}{(2\pi)^4}\Big[\mathcal{A}(B_c\to \eta_c D_{(s)})\Big]\Big[g_{T_{cc}D^*D}\varepsilon_{\beta}(p_2)\Big]\Big[-g_{ \eta_c \bar{D}D^*}(q_1+p_1)_{\nu}\Big]  \nonumber  \\  
&&\frac{1}{q^2_1-m^2_1}\frac{1}{q^2_2-m^2_2}\frac{-g^{\nu\beta}+q^\nu q^\beta /q^2}{q^2-m^2}\mathcal{F}(q^2,m^2),  \\ \nonumber
\mathcal{M}_{c,g,k}&=&i^2 \int\frac{d^4 q}{(2\pi)^4}\Big[\mathcal{A}_{\mu\nu}(B_c\to J/\psi D_{(s)}^{*})\Big]\Big[g_{T_{cc}D^*D}\varepsilon_{\beta}(p_2)\Big]\Big[-g_{ \psi \bar{D}D^*}\varepsilon^{\sigma\alpha\tau\nu}q_{1\sigma}p_{1\tau}\varepsilon_{\nu}(p_1)\Big]  \nonumber  \\  
&&\frac{-g^{\mu\alpha}+q^\mu_1 q^\alpha_1 /q^2_1}{q^2_1-m^2_1}\frac{-g^{\nu\beta}+q^\nu_2 q^\beta_2 /q^2_2}{q^2_2-m^2_2}\frac{1}{q^2-m^2}\mathcal{F}(q^2,m^2),  \\ \nonumber
\mathcal{M}_{c,g,k}&=&i^2 \int\frac{d^4 q}{(2\pi)^4}\Big[\mathcal{A}_{\mu}(B_c\to J/\psi D_{(s)})\Big]\Big[g_{T_{cc}D^*D}\varepsilon_{\beta}(p_2)\Big]   \\ \nonumber  &&  \Big[ -g_{\psi \bar{D}^*D^*} \varepsilon^{\sigma}(p_1) ( g_{\nu\sigma} (p_1-q)_{\alpha} + g_{\alpha\nu}(q_1+q)_{\sigma}- g_{\alpha\sigma}(p_1+q_1)_{\nu} ) \Big]  \nonumber  \\
&&\frac{-g^{\mu\alpha}+q^\mu_1 q^\alpha_1 /q^2_1}{q^2_1-m^2_1}\frac{1}{q^2_2-m^2_2}\frac{-g^{\nu\beta}+q^\nu q^\beta /q^2}{q^2-m^2}\mathcal{F}(q^2,m^2),  \\ \nonumber
\mathcal{M}_{d,h,l}&=&i^2 \int\frac{d^4 q}{(2\pi)^4}\Big[\mathcal{A}_{\nu}(B_c\to \eta_c D_{(s)}^*)\Big]\Big[g_{T_{cc}D^*D}\varepsilon_{\beta}(p_2)\Big]    \Big[ g_{\eta_c \bar{D}^*D} (q_1+q)_{\sigma} \varepsilon^{\sigma}(p_1) \Big]  \nonumber  \\
&&\frac{1}{q^2_1-m^2_1}\frac{-g^{\nu\beta}+q_2^\nu q_2^\beta /q_2^2}{q^2_2-m^2_2}\frac{1}{q^2-m^2}\mathcal{F}(q^2,m^2),  \\ \nonumber
\mathcal{M}_{d,h,l}&=&i^2 \int\frac{d^4 q}{(2\pi)^4}\Big[\mathcal{A}(B_c\to \eta_c D_{(s)})\Big]\Big[g_{T_{cc}D^*D}\varepsilon_{\beta}(p_2)\Big]    \Big[ g_{\eta_c \bar{D}^*D^*} \varepsilon_{\mu\nu\alpha\sigma} q^{\mu} p_1^{\alpha} \varepsilon^{\sigma}(p_1) \Big]  \nonumber  \\
&&\frac{1}{q^2_1-m^2_1}\frac{1}{q^2_2-m^2_2}\frac{-g^{\nu\beta}+q^\nu q^\beta /q^2}{q^2-m^2}\mathcal{F}(q^2,m^2),  \\ \nonumber
\end{eqnarray}
and the amplitudes of the decays $B_c \to T_{cc(s)}^* \bar{D}^{(*)}$ in Fig.~\ref{triangle2} are written as 
\begin{eqnarray}
\mathcal{M}_{m,q,u}&=&i^2 \int\frac{d^4 q}{(2\pi)^4}\Big[\mathcal{A}_{\mu\nu}(B_c\to J/\psi D_{(s)}^{*})\Big]\Big[g_{T_{cc}D^*D^*}\varepsilon_{\phi\theta\beta\omega }p_2^{\phi} \varepsilon^{\theta}(p_2)\Big]\Big[-g_{ \psi \bar{D}D^*}\varepsilon^{\sigma\alpha\tau\rho}q_{1\sigma}q_{\tau}\Big]  \nonumber  \\  
&&\frac{-g^{\mu\alpha}+q^\mu_1 q^\alpha_1 /q^2_1}{q^2_1-m^2_1}\frac{-g^{\nu\beta}+q^\nu_2 q^\beta_2 /q^2_2}{q^2_2-m^2_2}\frac{-g^{\rho\omega}+q^{\rho}q^{\omega}/q^2}{q^2-m^2}\mathcal{F}(q^2,m^2),  \\ \nonumber
\mathcal{M}_{n,r,v}&=&i^2 \int\frac{d^4 q}{(2\pi)^4}\Big[\mathcal{A}_{\nu}(B_c\to \eta_c D_{(s)}^{*})\Big]\Big[g_{T_{cc}D^*D^*}\varepsilon_{\phi\theta\beta\omega }p_2^{\phi} \varepsilon^{\theta}(p_2)\Big]\Big[-g_{ \eta_c \bar{D}D^*}(q_1+p_1)_{\rho}\Big]  \nonumber  \\  
&&\frac{1}{q^2_1-m^2_1}\frac{-g^{\nu\beta}+q^\nu_2 q^\beta_2 /q^2_2}{q^2_2-m^2_2}\frac{-g^{\rho\omega}+q^{\rho}q^{\omega}/q^2}{q^2-m^2}\mathcal{F}(q^2,m^2), 
\end{eqnarray}
\begin{eqnarray}
\mathcal{M}_{o,s,w}&=&i^2 \int\frac{d^4 q}{(2\pi)^4}\Big[\mathcal{A}_{\mu\nu}(B_c\to J/\psi D_{(s)}^{*})\Big]\Big[g_{T_{cc}D^*D^*}\varepsilon_{\phi\theta\beta\omega }p_2^{\phi} \varepsilon^{\theta}(p_2)\Big] 
 \\ \nonumber
 &&
  \Big[-g_{\psi \bar{D}^*D^*} \varepsilon^{\sigma}(p_1) ( g_{\alpha\sigma} (p_1-q)_{\rho} + g_{\rho\alpha}(q_1+q)_{\sigma}- g_{\rho\sigma}(p_1+q_1)_{\alpha} )\Big]  \nonumber  \\  
&&\frac{-g^{\mu\alpha}+q^\mu_1 q^\alpha_1 /q^2_1}{q^2_1-m^2_1}\frac{-g^{\nu\beta}+q^\nu_2 q^\beta_2 /q^2_2}{q^2_2-m^2_2}\frac{-g^{\rho\omega}+q^{\rho}q^{\omega}/q^2}{q^2-m^2}\mathcal{F}(q^2,m^2),  \\ \nonumber
\mathcal{M}_{p,t,x}&=&i^2 \int\frac{d^4 q}{(2\pi)^4}\Big[\mathcal{A}_{\nu}(B_c\to \eta_c D_{(s)}^{*})\Big]\Big[g_{T_{cc}D^*D^*}\varepsilon_{\phi\theta\beta\omega }p_2^{\phi} \varepsilon^{\theta}(p_2)\Big]\Big[g_{ \eta_c \bar{D}^*D^*}\varepsilon^{\sigma\alpha\tau\rho}p_{1\sigma}q_{\tau}\varepsilon_{\alpha}(p_1)\Big]  \nonumber  \\  
&&\frac{1}{q^2_1-m^2_1}\frac{-g^{\nu\beta}+q^\nu_2 q^\beta_2 /q^2_2}{q^2_2-m^2_2}\frac{-g^{\rho\omega}+q^{\rho}q^{\omega}/q^2}{q^2-m^2}\mathcal{F}(q^2,m^2), 
\end{eqnarray}

 With the  amplitudes of two-body decays, one can calculate  
 the corresponding partial decay widths  as
 \begin{eqnarray}
\Gamma=\frac{1}{2J+1}\frac{1}{8\pi}\frac{|\vec{p}|}{m^2}\bar{|\mathcal{M}|}^{2},
\end{eqnarray}
where $J$ is the total angular momentum of the initial state, the overline indicates the sum over the polarization vectors of final states, and $|\vec{p}|$ is the momentum of either final state in the rest frame of  initial state.

\section{Results and Discussion}

\begin{table}[!h]
\caption{Masses and quantum numbers of mesons relevant to the present work~\cite{ParticleDataGroup:2020ssz}. \label{mass}}
\begin{tabular}{ccc|ccc|ccc}
  \hline\hline
   Meson & $I (J^P)$ & M (MeV) &    Meson & $I (J^P)$ & M (MeV) &  Meson & $I (J^P)$ & M (MeV)   \\
  \hline  
      $\pi^{0}$ & $1(0^-)$ & $134.977$  &    $\pi^{\pm}$ & $1(0^-)$ &$139.570$ &    $\pi$ & $1(0^-)$ &$138.039$ \\
    $K^{0}$ & $\frac{1}{2}(0^-)$ & $497.611$  &    $K^{\pm}$ & $\frac{1}{2}(0^-)$ & $493.677$  & $K$ & $\frac{1}{2}(0^-)$ &$495.644$ \\
  $D^{0}$ & $\frac{1}{2}(0^-)$ & $1864.84$  &    $D^{\pm}$ & $\frac{1}{2}(0^-)$ & $1869.66$ & $D$ & $\frac{1}{2}(0^-)$ &$1867.25$ \\
  $D^{\ast0}$ & $\frac{1}{2}(1^-)$ & $2006.85$ &  $D^{\ast\pm}$ & $\frac{1}{2}(1^-)$ & $2010.26$ & $D^*$ & $\frac{1}{2}(1^-)$ &$2008.56$
  \\
     $D_s^{\pm}$ & $0(0^{-})$ & $1968.35$ & $D_s^{*\pm}$ & $0(1^{-})$ & $2112.2$  &  $T_{cc}$ & $0(1^{+})$ & $3874.74$  
     \\  $B^{\pm}$ & $\frac{1}{2}(0^-)$ & $5279.34$ &  $B^0$ & $\frac{1}{2}(0^-)$ & $5279.66$  &  $B_c$   & $0(0^-)$  & $6274.47$
  \\
  $\eta_c$ & $0(0^-)$ & $2984.1$ &  $J/\psi$ & $0(1^-)$ & $3096.90$  &  $B_c$   & $0(0^-)$  & $6274.47$
  \\
 \hline \hline
\end{tabular}
\label{tab:masses}
\end{table}

In Table~\ref{tab:masses}, we tabulate the masses and quantum numbers of relevant particles involved in this work.
In the heavy quark and SU(3)-flavor symmetry limits,  the contact-range potentials of  $I(J^P)=0(1^+)$~$DD^*$, $I(J^P)=0(1^+)$~$D^*D^*$, $I(J^P)=1/2(1^+)$~$D_sD^*$, and $I(J^P)=1/2(1^+)$~$D_s^*D^*$ systems are identical.  
By treating the doubly charmed tetraquark state $T_{cc}$ as a $DD^*$ bound state, we first determine the potential value and subsequently predict the masses of the other five states as shown in Table~\ref{massTcc}. Our results show that there exist five additional doubly charmed tetraquark states, which are slightly more bound than the $T_{cc}$ state. The majority of theoretical studies suggested the existence of an isoscalar $I(J^P)=0(1^+)$ $D^*D^*$ molecular bound state~\cite{Dong:2021bvy,Albaladejo:2021vln,Dai:2021vgf,Peng:2023lfw,Chen:2024bre,Whyte:2024ihh}. However, the existence of $I(J^P)=1/2(1^+)$~$D_sD^*$ and $I(J^P)=1/2(1^+)$~$D_s^*D^*$ bound states have large uncertainty~\cite{Dai:2021vgf,Peng:2023lfw,Chen:2024bre}\footnote{ A recent Lattice QCD study showed that the potentials of the $I(J^P)=1/2(1^+)$~$D_sD^*$ and $I(J^P)=1/2(1^+)$~$D_s^*D^*$ systems are too weak to form hadronic molecules~\cite{Shrimal:2025ues}.}.  If taking into account the uncertainty of SU(3)-flavor symmetry of $20\%$~\cite{Wu:2025fzx}, we find that the $I(J^P)=1/2(1^+)$~$D_sD^*$ and $I(J^P)=1/2(1^+)$~$D_s^*D^*$ systems are hardly to bind for their weaker potentials. The SU(3)-flavor symmetry-breaking effects for hadron-hadron interactions require further experimental confirmation. 
In the following, we assume that there exist six hadronic molecules as shown in Table~\ref{massTcc}, and then present a systematic investigation of their partial decay channels and production mechanism in $B_c$ decays, offering additional insights for experimental measurements.

\begin{table}[!h]
\centering
\caption{ Masses of $ {D}D^*$,  $ {D}^*D^*$,  $ {D}D_{s}^*$, and $ {D}^*D_s^*$ with $J^{P}=1^{+}$ for a cutoff of $\Lambda=1$ GeV. 
  }
\label{massTcc}
\begin{tabular}{cccccc}
\hline\hline
Molecule & $I$ & $J^{P}$  & $m$ (MeV)
& Experimental data (MeV)
    \\
  \hline
  $D {D}^{\ast}$ & $0$ & $1^{+}$
  & Input  & 3874.74  \\
  $D^{\ast}{D}^{\ast}$ & $0$ & $1^{+}$ & 4015.28  & -    \\
  \hline
  $D_{s}{D}^*$ & $\frac{1}{2}$ & $1^{+}$  & 3978.23/3974.55  & - \\
  $D_{s}^*{D}^*$ & $\frac{1}{2}$ & $1^{+}$    & 4119.98/4116.60  &-  
   \\
  \hline \hline
\end{tabular}
\end{table}

In Table~\ref{decayTcc}, we present the partial decay widths of the $DD^*$ and $D_sD^*$ molecules. Our results indicate that the partial decay widths   $T_{cc}^{+} \to D^0 D^0 \pi^+ $, $T_{cc}^{+} \to D^+ D^0 \pi^0 $, and $T_{cc}^{+} \to D^+ D^0 \gamma $ are around $23.69$, $19.82$, and $9.40$ keV, respectively.  The  total width of $T_{cc}$ is around $52.91$~keV, consistent with the experimental data~\cite{LHCb:2021auc}, which supports the molecular nature of $T_{cc}$ from the perspective of its decays.  For the SU(3)-flavor partners of  the $DD^*$ molecules, the total widths of $T_{ccs}^{+}$   and  $T_{ccs}^{++}$    are  around $20.91$ and $25.26$~keV, respectively,  almost half of the width of $T_{cc}$. The smaller widths of  $T_{ccs}^{+}$   and  $T_{ccs}^{++}$ are caused by the particularly weak couplings of $g_{D_{s}^*D_s\pi}$ and $g_{D_{s}^*D_s\gamma}$.     The $T_{ccs}^+$ molecule mainly decays into  $D_s^+D^0 \pi^0$  and $D_s^+D^0 \gamma$, which are difficult to observe by LHCb because of the low efficiency of observing  $\pi^0$ and $\gamma$. In addition to decaying into the above channels, the $T_{ccs}^{++}$ molecule dominantly decays into  $D_s^+D^0\pi^+$,  which is a favored channel for experimental observation with large branching fraction and high detection efficiency. Therefore, we suggest searching for the $T_{ccs}^{++}$ molecule in $D_s^+D^0\pi^+$ final states.

\begin{table}[!h]
\centering
\caption{ Partial decay widths  of $ {D}D^*$ and  $ {D}_sD^*$  molecules. 
  }
\label{decayTcc}
\begin{tabular}{c|ccccc}
\hline\hline
Molecule & $D^0 {D}^{0} \pi^+$ & $D^0 {D}^{+} \pi^0$  &  $D^0 {D}^{+} \gamma$    &  Total 
    \\
  $D {D}^{\ast}$(keV) & $23.69$ & $19.82$
  & $9.40$  & $52.91$ \\   \hline
  Molecule & $D_s^+ D^0  \pi^0$ & $D_s^+ {D}^{0} \gamma$  &     &  Total 
    \\
  $D_s {D}^{\ast}$(keV) & $8.00$ & $12.91$
  &   & 20.91 \\   \hline
  Molecule & $D_s^+ {D}^{0} \pi^+$   & $D_s^+ D^+  \pi^0$ &  $D_s^+ {D}^{0} \gamma$    &  Total 
    \\
  $D_s {D}^{\ast}$(keV) & $16.82$     & 7.40 & $1.04$
     & 25.26 \\
  \hline \hline
\end{tabular}
\end{table}

The two-body decay modes of the $T_{cc}^*$ molecule proceed via the triangle diagram mechanisms, where an unknown parameter $\alpha$ in the form factors is introduced.  Following the studies of the heavy hadron decays~\cite{Cheng:2004ru,Yu:2017zst} and the spectrum of hadronic molecular states~\cite{Liu:2024uxn},  we take the parameter $\alpha=1$ as input.  
In Table~\ref{decay1Tcc}, we present the partial decay widths of the  $ {D}^*D^*$  and $ {D}^*D_s^*$ molecules, whose widths are much larger than their HQSS partners since two-body decays are allowed. We find that the total width of the  $T_{cc}^*$ molecule is around $2.5$ MeV, consistent with the value calculated in  Ref.~\cite{Dai:2021vgf}. As shown in Table~\ref{decay1Tcc}, the three-body partial decay widths of  the  $T_{cc}^*$ molecule vary from several to tens keV, and the dominant three-body decay is $D^{*+}D^0\gamma$.  
Taking into account the branching fractions of $\mathcal{B}(T_{cc}\to D^0D^0\pi^+)=0.45$,   $\mathcal{B}(T_{cc}\to D^0D^+\pi^0)=0.37$, and $\mathcal{B}(T_{cc}\to D^0D^+\gamma)=0.18$, we can estimate the four-body partial widths of $T_{cc}^*$ molecule, i.e.,   $\Gamma(T_{cc}^*\to D^0D^0\pi^+\pi^0)=6.03$~keV, $\Gamma(T_{cc}^*\to D^0D^+\pi^0\pi^0)=4.96$~keV,   $\Gamma(T_{cc}^*\to D^0D^0\pi^+\gamma)=9.40$~keV, $\Gamma(T_{cc}^*\to D^0D^+\pi^0\gamma)=10.14$~keV, and $\Gamma(T_{cc}^*\to D^0D^+\gamma\gamma)=3.76$~keV.     
In Refs.~\cite{Jia:2022qwr,Jia:2023hvc}, the three-body and four-body pionic and radiative decays of the  $T_{cc}^*$ molecule were investigated; their results are a bit larger than ours. The branching fractions of the two-body decay $T_{cc}^*\to DD^*$ are around $96\%$, which indicates that  the promising channel of  experimental search for the $T_{cc}^*$ molecule are
$D^0D^{*+}$ and $D^+D^{*0}$ channels because the branching fractions of three-body and four-body decays are much smaller than those of two-body decays.

\begin{table}[!h]
\centering
\caption{ Partial decay widths  of   $ {D}^*D^*$  and $ {D}^*D_s^*$ molecules. 
  }
\label{decay1Tcc}
\begin{tabular}{c|ccccccccccccc}
\hline\hline
Molecule & $D^0 {D}^{*+}$ & $D^+ {D}^{*0} $  &  $ T_{cc} \pi^0$   &   $T_{cc} \gamma$   &   $D^{*+}D^0 \pi^0$    &   $D^{*0}D^0 \pi^+$  &   $D^{*0}D^+ \pi^0$  &$D^{*0}D^+ \gamma$  &$D^{*+}D^0 \gamma$   & Total 
    \\
  $D^* {D}^{\ast}$(keV) & $1215.76$ & $1215.76$ 
  & $13.41$  & $20.89$    & $10.12$    & $11.61$   & $4.84$   &$1.98$   & $31.64$  &  $2526.01$  \\   \hline
  Molecule &  $D^0 {D}_s^{*+}$  &  $D^{*0} {D}_s^{+}$   &  $T_{ccs}^+\pi^0$  &  $T_{ccs}^+\gamma$   &   $D_s^{*+} D^0  \pi^0$ & $D_s^+ {D}^{*0} \pi^0$  &    &  $D^{*0} {D}_s^{+} \gamma$      &      $D_s^{*+} {D}^{0} \gamma$   &  Total 
    \\
  $D_s^* {D}^{\ast}$(keV)   &  $694.18$  &  $656.66$   &  $27.55$  &$13.96$    &   $6.07$ & $0.001$  &   &  $0.09$   &  $29.85$  &  $1428.36$\\  \hline 
    Molecule &   $D^+ {D}_s^{*+}$  &  $D^{*+} {D}_s^{+}$   & $T_{ccs}^{++}\pi^0$   & $T_{ccs}^{++}\gamma$   &   $D_s^{*+} D^+  \pi^0$ & $D_s^{*+} {D}^{0} \pi^+$  &  $D^{*+} {D}_s^{+} \pi^0$ &     $D^{*+} {D}_s^{+} \gamma$      &      $D_s^{*+} {D}^{+} \gamma$   &  Total 
    \\
  $D_s^* {D}^{\ast}$(keV)   &  $687.78$   &  $654.28$   &  $20.19$  &  $1.12$   &   $2.36$ & $5.88$  &  $0.001$    &  $0.09$   & $1.85$  &$1373.55$ \\  
  \hline \hline
\end{tabular}
\end{table}

The  $D_s^*D^*$ and  $D^*D^*$ molecules have similar decay mechanisms, while the decay widths of the $D_s^*D^*$ molecules are around half of the  $D^*D^*$ molecule, as shown in Table~\ref{decay1Tcc}.  For the two-body decays, the FSIs  $D^*D^*\to DD^*$ and $D_s^*D^* \to D_sD^*/D_s^*D$ processes proceed by one pion exchange and one kaon exchange, respectively,  where the heavier kaon mass suppresses the two-body partial decay widths of the $D^*D_s^*$ molecules. Moreover, the violation of isospin conservation in the strong decay $D_s^* \to D_s \pi$, along with the narrow width of the $D_s^*$ state, naturally leads to small coupling constants for $g_{D_s^* D_s \pi}$ and $g_{D_s^* D_s \gamma}$. This, in turn, leads to very small partial widths for the $D_s^*D^*$ molecules decaying into  $D^*D_s\pi$ and  $D^*D_s\gamma$,  as shown in Table~\ref{decay1Tcc}. However, 
the width of the decay $T_{ccs}^{*+} \to D_{s}^{*+}D^0 \gamma$ is similar to that of $T_{cc}^* \to D^{*+}D^0\gamma$ due to the same decay mechanism, which proceed via the decay of $D^{*0} \to D^0 \gamma$. 
The  branching fractions   
$\mathcal{B}(T_{ccs}^{*+} \to D_sD^*/D_s^*D)=95\%$  and $\mathcal{B}(T_{ccs}^{*++} \to D_sD^*/D_s^*D)=98\%$ indicate that the $T_{ccs}^{*+}$ and $T_{ccs}^{*++}$ molecules dominantly decay into $D_sD^*/D_s^*D$, similar to the decays of $T_{cc}$. Therefore, the two-body channels are thus promising for experimental searches of the $D_s^*D^*$ molecules.

Unlike the decays of $B$ mesons, the experimental observation of $B_c$ meson decay modes remains very limited. In this work, we employ the naive factorization approach to predict the branching fractions of $B_c \to J/\psi(\eta_c) \bar{D}_s^{(*)}$ decays as shown in Table~\ref{productionTcc}~\cite{Bauer:1986bm}. We find that the branching fractions for $B_c \to J/\psi(\eta_c) D^{(*)}$ are typically of order $10^{-4}$, while those for $B_c \to J/\psi(\eta_c) D_s^{(*)}$ are about 20 times larger, such a ratio is consistent with $V_{cs}^2/V_{cd}^2\sim 18.5$. Then, we predict the branching fractions of the decays $B_c \to T_{cc(s)}^{*}\bar{D}_{(s)}^{(*)}$ as a function of the parameter $\alpha$ varying from $1$ to $2$ as shown in Table~\ref{productionTcc}.  
The branching fraction of the decay $B_c \to T_{cc}\bar{D}^0$ is $(0.82\text{–}2.07) \times 10^{-7}$, almost   fifty  times smaller than those of the decays $B_c \to T_{ccs}^+\bar{D}^{0}$  and  $B_c \to T_{ccs}^{++}{D}^{-}$, consistent with Ref.~\cite{Li:2023hpk}. Similarly,  the branching fractions of the decays of $B_c \to T_{ccs}\bar{D}^{*}$, $B_c \to T_{ccs}^*\bar{D}$, and $B_c \to T_{ccs}^*\bar{D}^{*}$   are two order of magnitude larger than those of the decays of  $B_c \to T_{cc}\bar{D}^{*}$, $B_c \to T_{cc}^*\bar{D}$, and $B_c \to T_{cc}^*\bar{D}^{*}$, respectively. Our findings indicate that the yields of doubly charmed tetraquarks with strangeness $S=-1$ are relatively high in $B_c$ decays, primarily due to the large value of the CKM matrix element $V_{cs}$.  The   branching fraction of the decays $B_c \to T_{cc(s)}\bar{D}^*$ are similar to those of the decays $B_c \to T_{cc(s)}^*\bar{D}$,  larger than those of the decays $B_c \to T_{cc(s)}^*\bar{D}^*$ and $B_c \to T_{cc(s)}\bar{D}$.

\begin{table}[!h]
\centering
\caption{ Branching fractions   of the decays $ B_{c} \to J/\psi \bar{D}_s^{(*)}$ and  $ B_{c} \to \eta_c \bar{D}_s^{(*)}$ as well as  the branching fractions of the  decays  $ B_{c} \to  T_{cc}^{(*)}\bar{D}^{(*)}$ and  $ B_{c} \to  T_{ccs}^{(*)}\bar{D}^{(*)}$.   
  }
\label{productionTcc}
\begin{tabular}{c|ccccc}
\hline\hline
Decay Channels & $\eta_c D^+$ & $\eta_c D^{*+}$  &   $J/\psi  D^+$   &  $J/\psi  D^{*+}$
    \\
  Branching fraction($10^{-4}$) & $2.30$ & $2.02$
  & $1.09$  & $4.92$ \\   \hline
Decay Channels & $\eta_c D_s^+$ & $\eta_c D_s^{*+}$  &   $J/\psi  D_s^+$   &  $J/\psi  D_s^{*+}$
    \\
  Branching fraction($10^{-3}$) & $5.91$ & $4.53$
  & $2.62$  & $12.34$ \\   \hline 
  Decay Channels & $T_{cc}\bar{D}^0$ & $T_{cc}\bar{D}^{*0}$  &  $T_{cc}^*\bar{D}^0$   &  $T_{cc}^*\bar{D}^{*0}$ 
    \\
  Branching fraction($10^{-7}$) & $0.82\sim2.07$ & $5.92\sim 14.18$ 
  & $8.48\sim20.61$ & $3.12\sim9.28$  \\   \hline 
  Decay Channels & $T_{ccs}^+\bar{D}^0$ & $T_{ccs}^+\bar{D}^{*0}$  &  $T_{ccs}^{*+}\bar{D}^0$   &  $T_{ccs}^{*+}\bar{D}^{*0}$ 
    \\
  Branching fraction($10^{-5}$) & $0.55\sim1.24$ & $1.52\sim 3.70$ 
  & $1.98\sim4.88$ & $0.56\sim1.80$  \\   \hline 
  Decay Channels & $T_{ccs}^{++}{D}^-$ & $T_{ccs}^{++}{D}^{*-}$  &  $T_{ccs}^{*++}{D}^-$   &  $T_{ccs}^{*++}{D}^{*-}$ 
    \\
  Branching fraction($10^{-5}$) & $0.53\sim1.18$ & $1.18\sim 2.87$ 
  & $1.87\sim4.62$ & $0.53\sim1.72$  \\  
  \hline \hline
\end{tabular}
\end{table}

For the experimental measurements, we must account for both the decay channels of the $T_{cc(s)}^{(*)}$ state and the detection efficiency of the experiment. Taking into account the branching fraction $\mathcal{B}(T_{cc} \to D^0D^0\pi^+) \sim 0.5$, we estimate the following branching fractions:
$\mathcal{B}[B_c \to (T_{cc} \to D^0D^0\pi^+)\bar{D}^0] \sim (0.4\text{–}1) \times 10^{-7}$ and
$\mathcal{B}[B_c \to (T_{cc} \to D^0D^0\pi^+)\bar{D}^{*0}] \sim (3\text{–}7) \times 10^{-7}$.
However, the latter decay channel is not recommended experimentally, as the $\bar{D}^{*0}$ meson must be reconstructed via $\bar{D}^0$ and $\pi^0$ or $\gamma$, and the $\pi^0$ meson and $\gamma$ photon have low detection efficiency at the LHCb. Therefore, we suggest observing the $T_{cc}$ state in the decay  $B_c \to   D^0D^0\pi^+ \bar{D}^0$.  Combining the branching fractions of   $\mathcal{B}(T_{cc}^*\to D^0D^{*+} )\sim 0.5$  and  $\mathcal{B}(T_{cc}^*\to D^{*0}D^{+} )\sim 0.5$, we obtain the branching fractions  $\mathcal{B}[B_c \to (T_{cc}^* \to D^0 D^{*+})\bar{D}^0] \sim (4\text{–}10) \times 10^{-7}$,  $\mathcal{B}[B_c \to (T_{cc}^* \to D^{*0} D^{+})\bar{D}^0] \sim (4\text{–}10) \times 10^{-7}$, $\mathcal{B}[B_c \to (T_{cc}^* \to D^{*0} D^{+})\bar{D}^{*0}] \sim (2\text{–}5) \times 10^{-7}$, and $\mathcal{B}[B_c \to (T_{cc}^* \to D^{*0} D^{+})\bar{D}^{*0}] \sim (2\text{–}5) \times 10^{-7}$. Similarly, neglecting the decay modes with the $D^{*0}$ and $\bar{D}^{*0}$  mesons, we suggest searching for the $T_{cc}^*$ in the decay $ B_c   \to D^0 D^{*+}\bar{D}^0 $. We  obtain the branching fraction $\mathcal{B}[B_c \to (T_{ccs}^+ \to D_s^+ D^{0} \gamma)\bar{D}^0] \sim (3\text{–}8) \times 10^{-6}$, $\mathcal{B}[B_c \to (T_{ccs}^+ \to D_s^+ D^{0} \gamma)\bar{D}^{*0}] \sim (1\text{–}2) \times 10^{-5}$,  $\mathcal{B}[B_c \to (T_{ccs}^{++} \to D_s^+ D^{0} \pi^+){D}^{-}] \sim (4\text{–}8) \times 10^{-6}$, and $\mathcal{B}[B_c \to (T_{ccs}^{++} \to D_s^+ D^{0} \pi^+){D}^{*-}] \sim (1\text{–}2) \times 10^{-5}$, and  suggest searching for the $T_{ccs}^{++}$ state in the decay  $B_c  \to D_s^+ D^{0} \pi^+{D}^{-}$
or the decay  $B_c  \to D_s^+ D^{0} \pi^+{D}^{*-}$.  However, the observation of $T_{ccs}^+$ in $B_c$ decays for the LHCb experiment is challenged due to the final states with both $\pi^0$ meson and $\gamma$ photon.  Along with similar analysis, we obtain the branching fraction  $\mathcal{B}[B_c \to (T_{ccs}^{*+} \to D_s^+ D^{*0})\bar{D}^0] \sim (1\text{–}2) \times 10^{-5}$, $\mathcal{B}[B_c \to (T_{ccs}^{*+} \to D_s^+ D^{*0})\bar{D}^{*0}] \sim (0.5\text{–}1) \times 10^{-5}$, 
$\mathcal{B}[B_c \to (T_{ccs}^{*++} \to D_s^+ D^{*+}){D}^-] \sim (1\text{–}2) \times 10^{-5}$, and $\mathcal{B}[B_c \to (T_{ccs}^{*++} \to D_s^+ D^{*+}){D}^{*-}] \sim (0.5\text{–}1) \times 10^{-5}$, which indicate the likely observation of $T_{ccs}^{*++}$ in the decay $B_c  \to D_s^+ D^{*+}{D}^-$.

Taking into account the branching fraction $\mathcal{B}(D^{*+}\to D^0\pi^+)=0.677$, we obtain the branching fractions 
$\mathcal{B}[B_c \to (T_{cc} \to D^0D^0\pi^+)\bar{D}^0] \sim (0.4\text{–}1) \times 10^{-7}$, $\mathcal{B}[B_c \to (T_{cc}^* \to D^0 D^{0} \pi^+)\bar{D}^0] \sim (3\text{–} 7) \times 10^{-7}$,   $\mathcal{B}[B_c \to (T_{ccs}^{++} \to D_s^+ D^{0} \pi^+){D}^{-}] \sim (4\text{–}8) \times 10^{-6}$, 
$\mathcal{B}[B_c \to (T_{ccs}^{*++} \to D_s^+ D^{0}\pi^+){D}^-] \sim (0.7\text{–}1.4) \times 10^{-5}$. Furthermore, the $D_{s}^+$ meson, $D^0$ meson, and $D^-$ meson are experimentally  reconstructed by the decays of  $D_{s}^+ \to K^+K^-\pi^+$, $D^0 \to K^- \pi^+$, and $ D^- \to  K^+ \pi^-\pi^-$. Therefore, with  the branching fractions  $\mathcal{B}(D_{s}^+ \to K^+K^-\pi^+)=0.05$, $\mathcal{B}(D^0 \to K^- \pi^+)=0.04$, and $\mathcal{B}(D^- \to  K^+ \pi^-\pi^-)=0.1$, we can estimate the production rates of  $T_{cc}$, $T_{cc}^*$, $T_{ccs}^{++}$, and $T_{ccs}^{*++}$ in $B_{c}$ decays are around $10^{-12}$, $10^{-11}$, $10^{-9}$, and $10^{-9}$, respectively. The cross sections of the $B_{c}$ meson is around $\sigma(B_c)\sim 100$~nb at the LHC~\cite{Chang:2003cr}, and the event number of the $B_{c}$ meson would be expected to be around $10^{10}$ at run 3 and $10^{11}$ at high-luminosity LHC(HL LHC)~\cite{Cerri:2018ypt}. Therefore, the $T_{ccs}^{++}$ and $T_{ccs}^{*++}$ states are expected to be observed in $B_c$ meson decays during LHC Run 3 and the HL-LHC, whereas the experimental detection of $T_{cc}$ and $T_{cc}^*$ remains challenging.

\section{Summary }
\label{sum}

The study of exotic states provides a pathway to understanding the non-perturbative strong interaction, thereby elucidating the mechanisms by which these states are formed.
Up to now, the only experimentally observed exotic state with doubly charmed quarks is the $T_{cc}$ discovered by the LHCb collaboration.  
Recent studies have indicated that the properties (mass and width) of the $T_{cc}$ state are consistent with it being a $DD^*$ hadronic molecule, where the non-perturbative interaction, i.e., hadron-hadron interaction, plays a dominant role. In other words, the discovery of $T_{cc}$ provides the chance to study the $DD^*$ interaction. In this work, we employed the contact-range effective field theory (EFT) approach to predict the interactions in SU(3)-flavor and heavy-quark spin symmetry partners of $DD^*$, along with their corresponding pole positions. To better support experimental measurements for doubly charmed tetraquark states, we further investigated their partial decay widths and production mechanism in $B_c$ decays.

Our results suggest the existence of three doubly charmed hadronic molecules, $T_{cc}$, $T_{ccs}^+$, and $T_{ccs}^{++}$,  which are linked by SU(3)-flavor symmetry, along with their HQSS partners, namely $T_{cc}^*$, $T_{ccs}^{*+}$, and $T_{ccs}^{*++}$. However, the existence of doubly charmed tetraquark states with strangeness is associated with a large uncertainty. Based on the molecular picture, the $T_{cc}$, $T_{ccs}^+$, and $T_{ccs}^{++}$ states decay predominantly into the three-body final states $D^0D^0\pi^+$/$D^0D^+\pi^0$, $D_s^+D^0\gamma$/$D_s^+D^0\pi^0$, and $D_s^+D^0\pi^+$/$D_s^+D^+\pi^0$, respectively. In contrast, the $T_{cc}^*$, $T_{ccs}^{*+}$, and $T_{ccs}^{*++}$ states decay mainly into the two-body final states $D^0D^{*+}$/$D^+D^{*0}$, $D^0D_s^{*+}$/$D_s^+D^{*0}$, and $D^+D_s^{*+}$/$D_s^+D^{*+}$, respectively. The production rates for $T_{cc}$ and $T_{ccs}$ in $B_c$ decays are around $10^{-7}$ and $10^{-5}$, respectively. For $T_{cc}^*$ and $T_{ccs}^*$, the corresponding values are around $10^{-6}$ and $10^{-5}$. Combining the detection efficiency of final states, we estimated the branching fractions of  $\mathcal{B}[B_c \to (T_{cc} \to D^0D^0\pi^+)\bar{D}^0]$, $\mathcal{B}[B_c \to (T_{cc}^* \to D^0 D^{0} \pi^+)\bar{D}^0] $,   $\mathcal{B}[B_c \to (T_{ccs}^{++} \to D_s^+ D^{0} \pi^+){D}^{-}] $, 
$\mathcal{B}[B_c \to (T_{ccs}^{*++} \to D_s^+ D^{0}\pi^+){D}^-]$ are around $10^{-12}$, $10^{-11}$, $10^{-9}$, and $10^{-9}$, respectively.
Based on the cross section of the $B_c$ meson at the LHC, a significant number of events is expected — around $10^{10}$ during Run 3 and $10^{11}$ at the HL-LHC. Therefore, our calculations suggest that the $T_{ccs}^{++}$ and $T_{ccs}^{*++}$ states could be observed in $B_c$ decays in LHC Run 3 and at the HL-LHC, while the experimental discovery of $T_{cc}$ and $T_{cc}^*$ remains challenging. 
Such investigations can test the molecular nature of the $T_{cc}$ and provide useful guidance for experimental searches for doubly charmed tetraquark states in $B_c$ decays.

\section{Acknowledgments}
 
 This work is partly supported by the National Key R\&D Program of China under Grant No. 2023YFA1606703 and the National Natural Science Foundation of China under Grant No. 12435007. M.Z.L acknowledges support from the National Natural Science Foundation of China under Grant No.~12575086. Yi Zhang acknowledges support from the National Natural Science Foundation of China under Grants No. 12347182 and the Project funded by China Postdoctoral Science Foundation
2023M740190.

\appendix

\bibliography{reference}

\end{document}